# A Holistic Power Optimization Approach for Microgrid Control Based on Deep Reinforcement Learning

Fulong Yao[a,*], Wanqing Zhao[a], Matthew Forshaw[a], Yang Song[b]

[a]*School of Computing, Newcastle University, Newcastle upon Tyne, NE4 5TG, UK*
[b]*School of Mechanical and Electrical Engineering and Automation, Shanghai University, Shanghai, 200444, China*

**Abstract**

Microgrid systems integrated with renewable energy sources (RES) and energy storage systems (ESS) have played a crucial role in providing more secure and reliable energy and deepening the penetration of renewables. However, optimizing the operational control of such an integrated energy system lacks the joint consideration of environmental, infrastructural and economic objectives, not to mention the need to factor in the uncertainties from both supply and demand. This paper presents a holistic data-driven power optimization approach based on deep reinforcement learning (DRL) for microgrid control, considering the multiple needs of decarbonization, sustainability and cost-efficiency. First, two control schemes, namely the prediction-based (PB) and prediction-free (PF) schemes, are devised to formulate the control problem for a microgrid integrated with RES and ESS. Second, a holistic objective (reward) function is designed to account for the operational costs, carbon emissions, peak load and battery degradation of the microgrid system. Third, we develop a data-driven Double Dueling Deep Q Network (D3QN) architecture to optimize the power flows (particularly the charging and discharging behaviors of ESS) for real-time energy management. Finally, extensive simulations are conducted on a US microgrid energy system to demonstrate the effectiveness and superiority of the proposed approach. The results and analysis also suggest the respective circumstances for using the two control schemes when there are uncertainties in practical implementations.

## 1. INTRODUCTION

With the continuous surge in energy demand and the exacerbation of energy shortages, the microgrid landscape is undergoing a transformative shift towards a more clean and sustainable paradigm [1]. At present, a microgrid that integrates RES and ESS is playing a crucial role in the energy system transition due to its advantages in security, reliability and cleanliness [2]. RES such as photovoltaic (PV) and wind turbine (WT) help reduce the microgrid's dependence on fossil fuels or the national grid, thereby reducing energy costs and greenhouse gas emissions. This, however, exacerbates the uncertainty and uncontrollability of the integrated energy system [3].

*Corresponding author: Fulong Yao
  *Email addresses:* `f.yao3@newcastle.ac.uk` (Fulong Yao), `wanqing.zhao@newcastle.ac.uk` (Wanqing Zhao), `matthew.forshaw@newcastle.ac.uk` (Matthew Forshaw), `y_song@shu.edu.cn` (Yang Song)

The intermittent RES generation combined with the uncertain power demands (due to changing weather conditions and users' behaviours) poses a fundamental challenge to the power optimization of a microgrid system involving RES and ESS.

**Abbreviations**

| | |
|---|---|
| RES | renewable energy sources |
| ESS | energy storage systems |
| PV | photovoltaic |
| RL | reinforcement learning |
| DRL | deep reinforcement learning |
| MDP | markov decision process |
| SGD | stochastic gradient descent |
| FC | fully connected |
| TD | temporal difference |
| WT | wind turbine |

**Environment Symbols**

| | |
|---|---|
| $P_{g,t}$ | imported power from the national grid (kW) |
| $P_{r,t}$ | RES generation (kW) |
| $P_g^{max}$ | maximum imported power from the grid (kW) |
| $P_{b,t}$ | power flowing into/out battery (kW) |
| $P_{d,t}$ | aggregated power demand (kW) |
| $P_{u,t}$ | unmet power (kW) |
| $P_{b,sb}$ | constant standby loss of battery (kW) |
| $PR_t$ | electricity price ($/kWh) |
| $CI_t$ | carbon intensity (gCO2/kWh) |
| $E_{b,t}$ | state of charge (kWh) |
| $\eta_b$ | charge/discharge efficiency |
| $\triangle t$ | time interval (h) |
| $t$ | time instant |
| $E_b^{min}, E_b^{max}$ | maximum and minimum state of charge (kWh) |
| $E_{max}$ | maximum energy charged or discharged within $\triangle t$ |

**Reinforcement Learning Symbols**

| | |
|---|---|
| $S$ | state space |
| $A$ | action space |
| $R, R'$ | immediate reward |
| $P_M$ | state transition probability |
| $s, s'$ | state |
| $a, a'$ | action |
| $T$ | time horizon of state |
| $L(w)$ | Mean Squared Error |
| $\mathcal{A}(s,a)$ | advantage function |
| $\mathcal{V}(s)$ | state-value function |
| $G_t$ | discounted long-term reward |
| $\pi^*, Q_\pi^*$ | optimal policy and optimal action-value function |

**Parameters**

| | |
|---|---|
| $M$ | updating frequency for Target Network |
| $\epsilon$ | random exploration rate |
| $\gamma$ | discount factor |
| $K$ | length of time series sequences in each episode |
| $\alpha$ | coefficient of carbon emissions |
| $\beta$ | coefficient of peak load limit |
| $\lambda$ | coefficient of battery degradation |
| $w$ | weight of Q Network |
| $w^-$ | weight of Target Network |
| $lr$ | learning rate |
| $\tau$ | updating rate of Target Network |

To address the above challenge, control-oriented power optimization was thus studied for a more resilient and sustainable energy future [4, 5, 6]. For example, considering the battery lifetime characteristics, Das and Ni [7] devised a dynamic programming approach to optimize the microgrid power flows for reduced operational costs and extended battery life. Morstyn et al. [8] developed a novel model predictive control (MPC) approach to optimize the power flow among the distributed batteries such that the supply and demand are balanced and the operational costs are minimised. In addition, Sachs and Sawodny [9] designed a two-layer MPC method (the first layer optimizes energy dispatch between PV generation and power loads, while the second layer controls the power output from a diesel generator) for a hybrid microgrid system to minimize operational costs and enhance the system robustness against uncertainties from both supply and demand sides. All these approaches have demonstrated impressive performance in optimizing the power flows in a microgrid. However, as the microgrid systems and their operational requirements are getting more complex, such as the increase in the number of system variables and optimization objectives, these traditional approaches became incapable of expressing and effectively addressing the control optimization problems.

Recently, reinforcement learning (RL) has shown remarkable efficacy in optimizing the power management of complex microgrid systems, attributed to its powerful model-free and self-learning capabilities [10]. For example, Ojand and Dagdougui [11] developed a two-stage MPC integrating a RL model (with a discrete state space) to control the distributed energy resources and power demand for the reduction of operational costs and enhancement of operating efficiency. However, a RL of this kind suffers from the curse of dimensionality, where the number of states grows exponentially with the increase of state variables. Deep reinforcement learning (DRL) overcomes this weakness by employing deep neural networks to approximate the value functions for microgrid control, allowing it to handle larger and more complex microgrid environments with improved performance and efficiency [12]. For instance, Alabdullah and Abido [13] developed a deep Q network (DQN) approach, which considers the stochastic behaviors of supply & demand profiles and pricing signals, to schedule the system operations and reduce costs. Yu et al. [14] further employed a double deep Q network (DDQN) for controlling energy storage systems, aiming to improve utilization of multiple renewable sources including PV, WT and hydrogen. These studies have demonstrated the vast potential and advantages of DRL for energy optimization in microgrids.

Moreover, the use of continuous state space in DRL, which accounts for the uncertainty of microgrid variables (e.g., load and demand) [14], has also been studied for microgrid control. For example, Cao et al. [15] proposed a data-driven Noisy Net-Double Deep Q Network (NN-DDQN), which constructed a state space containing near-future electricity price pattern and depth of charge as continuous variables, to optimize the battery energy arbitrage via load shifting. A data-driven online DRL method was devised in [16] to continuously control the microgrid using the past data (including dynamic power load, generations and electricity prices) and current depths of battery. Harrold et al. [17] developed a DRL-based MPC algorithm that used the predicted future data (including uncertain electricity price, demand, PV and WT generations) to maximize the utilization of RES generations and reduce operational costs. These studies have shown that either historical or predicted future data can be incorporated in state space to address complex control problems and yield favorable outcomes. However, there is an open question as to whether future or past data is more beneficial when defining the microgrid state space for DRL.

Furthermore, the goal of balancing between economic, environmental and infrastructural considerations has now posed another formidable challenge for power optimization in microgrid systems [18]. Optimizing the power management of microgrid systems only by relying on conventional considerations (such as operational cost) is no longer sufficient to meet our decarbonization and sustainability requirements. Fortunately, recent research has begun to consider optimizing the power management from these two aspects. For example, Rangel et al. [19] designed an optimization model based on the formulation of a mixed-integer linear programming (MILP) to balance the fuel consumption, carbon emissions and operational costs of a Diesel-PV-Battery microgrid. A cooperative operation model based on asymmetric nash bargaining was also developed in [20] for thermal-power microgrids to minimize carbon emissions and power sharing operational costs. Moreover, Wang et al. [21] proposed a low-carbon optimal scheduling strategy for district microgrids. It employed a novel carbon emission measuring technique based on the flow analysis theory to calculate the power consumed by microgrids. Though promising results were obtained, current research efforts are rather fragmented and remains in its early stages. Little has been done to systematically study the balance between the requirements of cost-efficiency, decarbonization and sustainability from the perspectives of economic, environmental and infrastructural considerations.

To address the complex interplay between decarbonization, sustainability and cost-efficiency, while factoring in the uncertainties from both the supply and demand, this paper proposes a holistic power optimization approach for microgrid control based on DRL. To this end, the main contributions can be summarized as:

1) A unified DRL framework is designed for microgrid control incorporating the influence of external and internal variables, in which a new objective (reward) function is devised to balance decarbonization, sustainability and cost-efficiency performance of the microgrid system;

2) A data-driven Double Dueling Deep Q Network (D3QN) architecture is developed to optimize the power management of microgrid systems integrated with RES and ESS;

3) Two control schemes (i.e., prediction-based and prediction-free) are devised within the unified DRL framework to leverage the learning competencies of D3QN for operating energy storages in different circumstances;

4) Extensive simulations are conducted to demonstrate the effectiveness of the proposed approach and suggest the different circumstances for using the two control schemes in practical implementations when uncertainties arise.

The subsequent sections of this paper are structured as follows. Section 2 introduces the model of the microgrid with RES and ESS. Section 3 provides an overview of classic DRL algorithms. Section 4 presents the proposed power optimization approach based on DRL, including a unified framework, a D3QN architecture and two data-driven control schemes. Simulation analysis and performance evaluation are given in Section 5, while conclusion is provided in Section 6.

## 2. MICROGRID SYSTEM AND ITS MODELING

This paper considers the control-oriented power optimization for a large class of microgrid systems that integrate RES and ESS to meet the energy demand of a local community. The schematic of such a microgrid is depicted in Figure 1, in which the arrows represent the direction of power, control and data flows. Here, energy can be supplied either from the national grid or RES (e.g., PV and WT), whilst the load is the total electricity demand of the locality. Then, battery is employed as a buffer to temporarily store renewable generations when there are surpluses after use, or to import electricity from the grid when energy is cheap and clean. Therefore, an additional supply of energy (which is low-cost and low-carbon) can be provided from battery for later use when there would be otherwise a need to import electricity from grid with a much higher price or carbon intensity.

To model the above microgrid, the power balance problem between the supply and load can

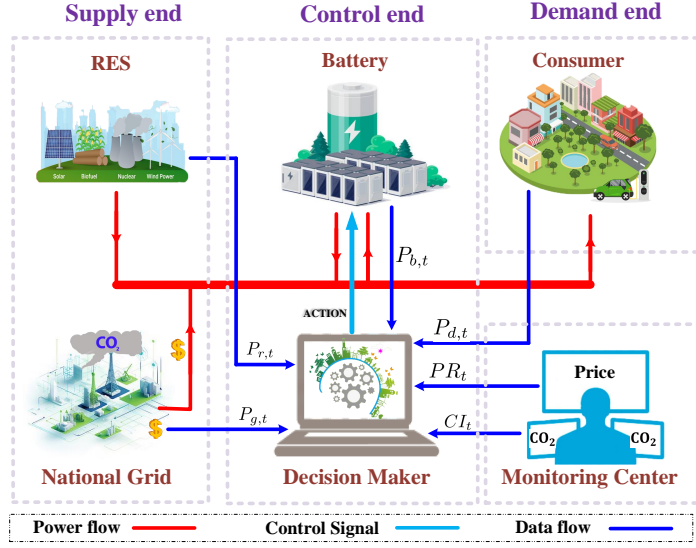

Figure 1: The schematic of the microgrid

be first expressed as:
$$P_{g,t} + P_{b,t} + P_{r,t} - P_{d,t} = 0 \tag{1}$$

where $t$ represents the time instant, $P_{g,t}$ (kW) is the power imported from the national grid, $P_{b,t}$ (kW) is the power flowing into/out the battery, $P_{r,t}$ (kW) is the power generated by RES and $P_{d,t}$ (kW) is the aggregated consumer demand. Meanwhile, the following constraints should be met:

$$P_{g,t} \geq 0 \tag{2}$$

$$P_{r,t} \geq 0 \tag{3}$$

$$P_{d,t} \geq 0 \tag{4}$$

$$\begin{cases} P_{b,t} > 0, & if\ discharge; \\ P_{b,t} = 0, & if\ idle; \\ P_{b,t} < 0, & if\ charge. \end{cases} \tag{5}$$

Then, we define the unmet power $P_{u,t}$ (kW) as:
$$P_{u,t} = P_{d,t} - P_{r,t} \tag{6}$$

By substituting (6) into (1), this gives

$$P_{g,t} + P_{b,t} - P_{u,t} = 0 \tag{7}$$

Assuming the charge/discharge efficiency of battery is $\eta_b \in (0,1)$ and there are no other losses, the following holds:

$$\begin{cases} 0 < P_{b,t} \cdot \triangle t \leq (E_{b,t} - E_b^{min}) \cdot \eta_b, & if\ discharge; \\ P_{b,t} = 0, & if\ idle; \\ \dfrac{E_{b,t} - E_b^{max}}{\eta_b} \leq P_{b,t} \cdot \triangle t < 0, & if\ charge. \end{cases} \tag{8}$$

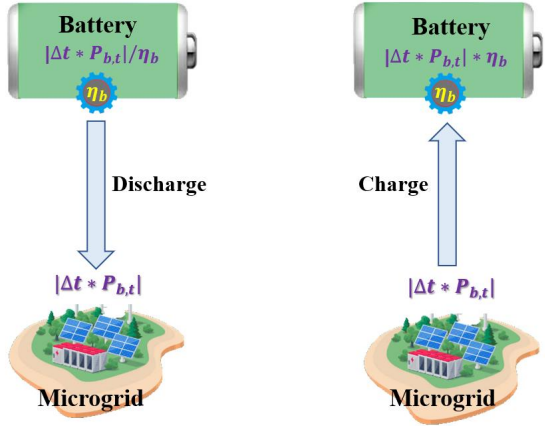

Figure 2: Battery discharging and charging processes

where $P_{b,t} \cdot \triangle t \cdot \eta_b$ (kWh) is the amount of charged energy within an interval $\triangle t$ (here, $P_{b,t} \cdot \triangle t / \eta_b$ being the discharged energy), $E_{b,t}$ (kWh) is the current state of charge (SOC), $E_b^{max}$ (kWh) and $E_b^{min}$ (kWh) represent the maximum and minimum state of charge, respectively. The processes of charging and discharging are illustrated in Figure 2.

Therefore, the state transition of the battery is expressed as [22]:

$$E_{b,t+1} = \begin{cases} E_{b,t} - (\triangle t \cdot P_{b,t})/\eta_b, & if \ P_{b,t} > 0; \\ E_{b,t} - \triangle t \cdot P_{b,sb}, & if \ P_{b,t} = 0; \\ E_{b,t} - \triangle t \cdot P_{b,t} \cdot \eta_b, & if \ P_{b,t} < 0. \end{cases} \quad (9)$$

$$P_{b,sb} \geq 0 \quad (10)$$

where $P_{b,sb} \geq 0$ (kW) indicates a constant standby loss rate of the battery.

Additionally, to prevent potential damage to the grid, the long-term stable operation of the microgrid requires the imported power to meet the peak load constraint as much as possible:

$$P_{g,t} \leq P_g^{max} \quad (11)$$

where $P_g^{max}$ (kW) represents the pre-set maximum peak load.

By substituting (7) into (11), the following is obtained:

$$P_{u,t} - P_{b,t} - P_g^{max} \leq 0 \quad (12)$$

The microgrid control system is required to make optimal decisions about when and how much to import power from the grid based on a number of variables, such as RES generation, unmet power, SOC of battery, electricity price and carbon intensity. Specifically, to achieve cost efficiency, decarbonization and sustainability, the microgrid is expected to:

1) buy as much power as possible and store it in the battery when the electricity price and carbon intensity are low;

2) give priority to use RES generations and the backup power in the battery to meet the demand when the price and carbon intensity are high;

3) reduce the peak load and slow down battery degradation to ensure long-term security and sustainability of the microgrid.

## 3. DEEP REINFORCEMENT LEARNING

### 3.1. Deep Q Network

Reinforcement learning (RL) is a machine learning paradigm where agents learn to make sequential decisions by interacting with an environment [23]. Despite its effectiveness in various applications, RL faces challenges in handling high-dimensional input spaces and addressing exploration-exploitation dilemma. Deep Q Network (DQN), developed by Google DeepMind [24], is a typical deep reinforcement learning (DRL) approach that combines deep neural networks with Q-learning (a typical RL algorithm) [25]. It addresses the limitations of traditional RL in discretizing environments involving high-dimensional state spaces. Generally, it uses a deep neural network to learn a control policy, which maps the states of the environment to optimal actions that maximize the expected cumulative reward over time [26]. In detail, the interaction between an agent and the environment can be modeled by a Markov Decision Process (MDP) with discrete time steps. This can be described by a tuple $\{S, A, R, P_M, \gamma\}$ [27], where $S$ represents the state space, $A$ is the action space, $R$ is the immediate reward, $P_M(s_{t+1}|s_t, a_t)$ is the state transition probability which shows the probability that the agent would move to state $s_{t+1}$ if it takes action $a_t \in A$ in state $s_t \in S$, and $\gamma$ is a discount factor that relates to the future reward.

Unlike the traditional Q-learning (which uses a table to store the expected values of the reward for each state-action pair), DQN adopts a deep neural network rather than a table to approximate the optimal value function $Q^*(s_t, a_t)$, as shown below:

$$Q(s_t, a_t; w_d) \approx Q^*(s_t, a_t) \qquad (13)$$

where $w_d$ is the network weights and $s_t$ and $a_t$ are the network inputs.

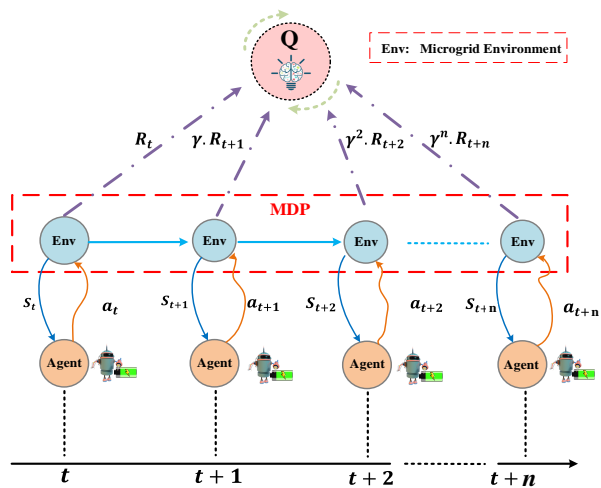

Figure 3: The interaction process between the environment and the agent

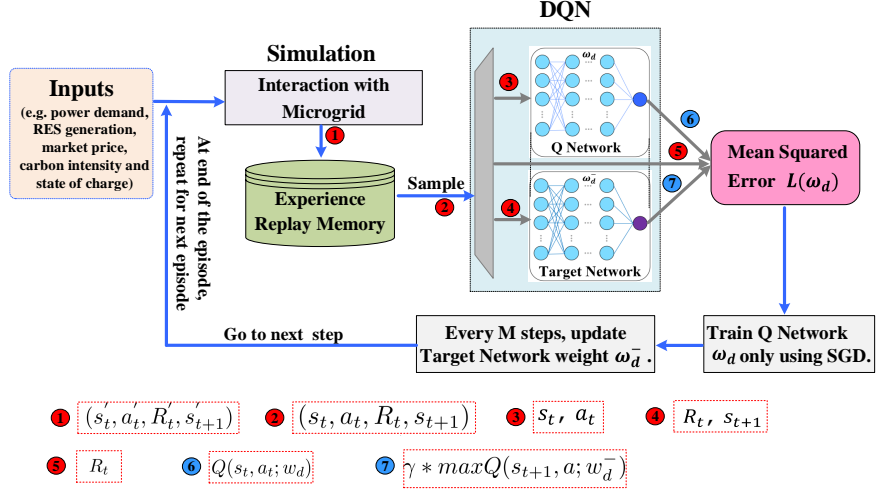

Figure 4: Flowchart of the learning process of DQN

The optimal value function can be learnt through a number of successive interactions between the agent and environment. At each time step (interaction), $Q(s_t, a_t)$ can be updated by a Bellman Equation [28, 29]:

$$Q(s_t, a_t) \leftarrow Q(s_t, a_t) + lr \cdot [R_t + \gamma \cdot \max_a Q(s_{t+1}, a) - Q(s_t, a_t)] \tag{14}$$

where $lr$ is the learning rate, $\gamma$ is the discount factor used to balance the importance between the immediate reward ($R_t$) and the potential future reward ($\max_a Q(s_{t+1}, a)$) of an action. A higher value of $\gamma$ means that future rewards are more certain and valuable. Here, $R_t + \gamma \cdot \max_a Q(s_{t+1}, a)$ is also called the TD target. The whole interaction process between a microgrid environment and the agent can be illustrated in Figure 3.

The flowchart of the learning process of DQN is shown in Figure 4. The objective is to minimize the Mean Squared Error $L(w_d)$ between the network output $Q(s_t, a_t; w_d)$ and TD target $R_t + \gamma \cdot \max_a Q(s_{t+1}, a; w_d^-)$:

$$L(w_d) = (R_t + \gamma \cdot \max_a Q(s_{t+1}, a; w_d^-) - Q(s_t, a_t; w_d))^2 \tag{15}$$

Here, a separate Target Network with weights $w_d^-$ is used to estimate the TD target value [30]. Generally, the Target Network weight $w_d^-$ is temporarily fixed while Q Network is under training with stochastic gradient descent (SGD) [31], but it will be then periodically updated based on the Q Network weight $w_d$ (i.e., directly copy $w_d$ to $w_d^-$ every $M$ steps). Moreover, an experience replay memory is employed to store the interaction experiences $(s_t', a_t', R_t', s_{t+1}')$ with the microgrid environment so that the agent can then randomly sample mini-batches from it to update the network weights. In this way, DQN allows the agent to reuse past experiences and breaks the sequential correlations, enhancing the diversity and efficiency of learning [32].

3.2. Variants of DQN

1) *Double Deep Q Network (DDQN)*: DDQN extends the DQN algorithm and aims to address the issue of overestimation of Q-values [33]. In traditional DQN, a network is used for both estimating the expected values and selecting actions. This can easily lead to overestimation of Q-values, especially in situations where there are multiple highly ranked actions. DDQN has proven to be an effective way to alleviate the overestimation issue by introducing a new DQN network to separate the value estimation from action selection. Generally, DDQN adopts one DQN with weight $w_{dd}$ to select the best action at current state and then uses the other DQN with weight $w_{dd}^-$ to estimate the target Q value based on the selected action. The objective of DDQN can be expressed as [33],

$$L(w_{dd}) = (R_t + \gamma \cdot Q(s_{t+1}, \underset{a}{argmax} Q(s_{t+1}, a; w_{dd}); w_{dd}^-) - Q(s_t, a_t; w_{dd}))^2 \tag{16}$$

2) *Dueling DQN*: Dueling DQN, introduced by Wang et al [34], is another variant of DQN that focuses on improving the learning and representation of action-value functions. In traditional DQN, the Q-values are directly estimated for each action in a given state. However, in many cases, different actions may have similar values and it is unnecessary to estimate their values individually [35]. Dueling DQN addresses this issue by decomposing the Q-value into two components: a state-value function $\mathcal{V}(s)$ that represents the expected value of being in a particular state (regardless of the action taken) and an advantage function $\mathcal{A}(s, a)$ that shows the relative significance of each action over other actions. The state-value function is responsible for estimating the baseline value for each state, while the advantage function quantifies the additional value offered by taking a particular action.

Rather than relying on one stream of fully connected (FC) layers to estimate the Q-values, the dueling DQN employs two streams of FC layers to calculate the value function $\mathcal{V}(s)$ and advantage function $\mathcal{A}(s, a)$, respectively. These two streams are then combined to produce a Q-value for each action, which can be represented as [34]:

$$Q(s, a) = \mathcal{V}(s) + \left( \mathcal{A}(s, a) - \frac{1}{|A|} \sum_{a' \in A} \mathcal{A}(s, a') \right) \tag{17}$$

## 4. D3QN-BASED OPTIMIZATION APPROACH

Considering the large number of states in the continuous space for microgrid modeling, this paper proposes a holistic data-driven power optimization approach based on DRL for controlling the battery charging and discharging behaviors. Figure 5 shows the framework of the proposed approach, in which two control schemes are devised to inference the control strategies using different state variables. Then, a data-driven D3QN architecture is developed to optimize the power flow for real-time energy management.

### 4.1. DRL Definition

In this paper, the main goal for the agent is to control the behavior of battery to reduce operational costs and carbon emissions as well as improving sustainability of the microgrid. First, the following microgrid environment with discrete actions and continuous states is defined.

1) *Action space*: The action space is discretized as (18) in which the behaviors $a_t \in A$ can be maximum charge (-1), half charge (-0.5), idle (0), half discharge (0.5) and maximum discharge (1), respectively.

$$A = \{-1, -0.5, 0, 0.5, 1\} \tag{18}$$

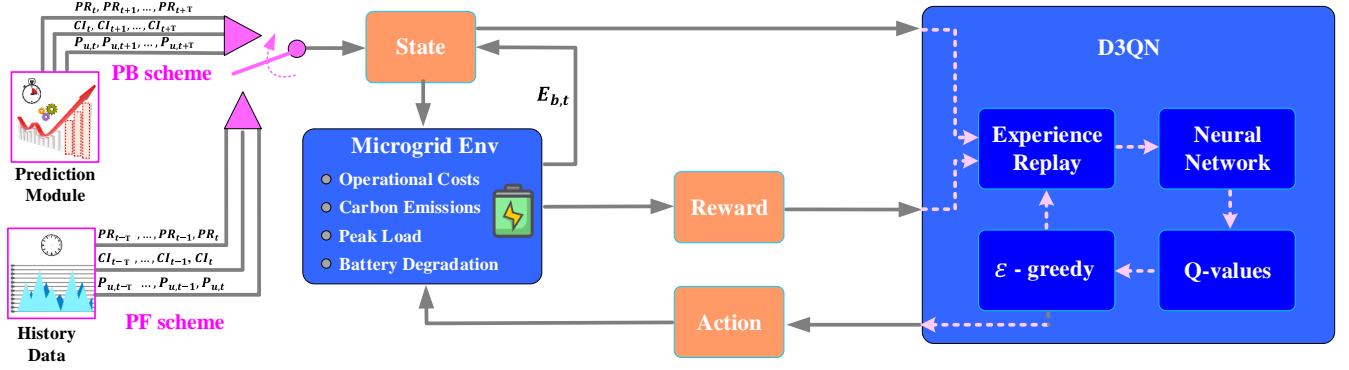

Figure 5: The framework of the proposed D3QN-based approach

These actions are relative values to the maximum amount of electricity ($E_{max}$ (kWh)) that can be transported to/from battery within an interval $\triangle t$. Thus, the resultant charged/discharged power can be computed by (19). It should be noted that $P_{b,t}$ is also required to meet the constraint in (8).

$$P_{b,t} = \frac{a_t \cdot E_{max}}{\triangle t} \tag{19}$$

$$E_{max} \geq 0 \tag{20}$$

2) *State space*: The continuous state space consists of two parts: the external variables (including electricity price, carbon intensity and unmet power) that are not controlled by the agent; and the internal variable (SOC of battery) that can be controlled via the aforementioned actions. In this paper, two control schemes (i.e., prediction-based and prediction-free) are employed to formulate the control problem, which are explained below.

① Prediction-based (PB) scheme: We employ the MPC concept where the predicted near-future statuses of the external variables are used in combination with battery status ($E_{b,t}$) to form the state space, as expressed in (21). The idea is to find the optimal control strategy for the current time step but taking into account future changes to the system. It should be noted that this paper only focuses on the controller design, while the prediction itself is beyond the scope of this paper.

$$s_t = (PR_t, PR_{t+1}, \ldots, PR_{t+T}, CI_t, CI_{t+1}, \ldots, CI_{t+T}, P_{u,t}, P_{u,t+1}, \ldots, P_{u,t+T}, E_{b,t}) \tag{21}$$

Here, $T$ indicates the time horizon to be considered for each variable, $PR_t$ is the electricity price ($/kWh), and $CI_t$ (gCO2/kWh) is the carbon intensity.

② Prediction-free (PF) scheme: As the name suggests, this scheme doesn't require a prediction model. Instead, it replaces predicted values with historical records for the external inputs in the state space. The idea is to implicitly capture the trend of variations in the external variables and incorporate them into the inference process for calculation of control actions. In this case, the state space is defined as (22). This scheme can effectively reduce the complexity of the control solution and avoid prediction uncertainties.

$$s_t = (PR_{t-T}, \ldots, PR_{t-1}, PR_t, CI_{t-T}, \ldots, CI_{t-1}, CI_t, P_{u,t-T}, \ldots, P_{u,t-1}, P_{u,t}, E_{b,t}) \quad (22)$$

It can be seen from (21) and (22) that merging the demand $P_{d,t}$ and RES generation $P_{r,t}$ together as the unmet power $P_{u,t}$ (as shown in (6)) greatly reduces the input dimension, from a space of $4(T+1)+1$ to $3(T+1)+1$.

3) *Reward function* : Given the main goal of this work, the reward function should reflect not only operational costs and carbon emissions, but also the reliability and sustainability of the microgrid. The immediate reward $R_t$ given by an action at time step $t$ is defined as follows to account for operational cost, carbon emission, peak load and battery degradation, respectively.

$$\begin{aligned} R_t &= PR_t \cdot P_{b,t} \cdot \triangle t + \alpha \cdot CI_t \cdot P_{b,t} \cdot \triangle t \\ &+ \beta \cdot min[0, P_{b,t} + P_g^{max} - P_{u,t}] \cdot \triangle t - \lambda \cdot |a_t| \cdot E_{max} \end{aligned} \quad (23)$$

where $\alpha \geq 0, \beta \geq 0, \lambda \geq 0$ are the weights controlling the importance of each objective. It should be noted that $PR_t$ and $CI_t$ should be normalized to same scale (e.g., [0,1]).

Here, we can see that the first two items represent the reductions in operational costs and carbon emissions resulting from adopting an appropriate action $P_{b,t}$ (compared to when the battery is not in use). When the RES generation $P_{r,t}$ cannot meet the demand $P_{d,t}$, the more the battery discharges ($P_{b,t} > 0$) at high price and carbon intensity (or the more the battery charges ($P_{b,t} < 0$) at low price and carbon intensity), the less the system produces operational costs and carbon footprints [15]. Correspondingly, $R_t$ becomes bigger. The third item indicates the penalty for breaking the peak load constraint. The use of mathematical operators ($min$) guarantees that the penalty is applied only when the instantaneous imported power exceeds the maximum allowed. The greater the excess, the heavier the penalty. The last item is related to the battery degradation cost which is simplified as a linear function proportional to charging and discharging actions [15]. The more the charging and discharging cycles, the more severe the battery degradation.

In addition, the weights in the reward function need to be carefully chosen based on practical considerations. In this paper, we considered the objectives in the following order: (i) the sustainable operation of the microgrid, (ii) operational cost reduction, (iii) carbon emission reduction and (iv) extended battery life. In particular, as battery degradation is a cumulative process and a single charge/discharge event has only a small effect on it. $\lambda$ should be chosen as a small value in the reward calculation. Given these considerations, the specification of coefficients in the reward function should follow:

$$\beta > 1 > \alpha \gg \lambda \geq 0 \quad (24)$$

*4.2. D3QN-based Approach*

A data-driven Double Dueling DQN (D3QN) architecture is developed in this paper, which combines the advantages of DDQN and Dueling DQN together to optimize microgrid controls. The architecture of this D3QN is shown in Figure 6. In the forward propagation, a mini-batch of experience tuple $(s_t, a_t, R_t, s_{t+1})$ is sampled from the experience replay memory. Then, the present state $s_t$ and action $a_t$ are passed to the Q Network to estimate the expected total reward $Q(s_t, a_t; w)$, while the next state $s_{t+1}$ is passed to a Target Network to calculate the future reward $Q(s_{t+1}, *; w^-)$. Noted that the target network is simply a copy of the Q network but updated every $M$ time steps to ensure the stability of the learning process. Based on the TD algorithm [28], to update the Q network, the loss $L(w)$ can be computed based on the difference between $Q(s_t, at; w)$ and $\gamma \cdot Q(s_{t+1}, *; w^-) + R_t$ (the latter is the sum of future and immediate rewards from the underlying action). In the backward propagation, $L(w)$ is then minimised by updating the weights $w$ in the Q Network. Moreover, the weights $w^-$ of the Target Network will only get updated every $M$ time steps using the weights $w$ from the Q network. Differing from the traditional DQN where the $w^-$ is adjusted by hard update (directly copying $w$ as $w^-$), this paper used the soft update (25) to adjust the target weights $w^-$.

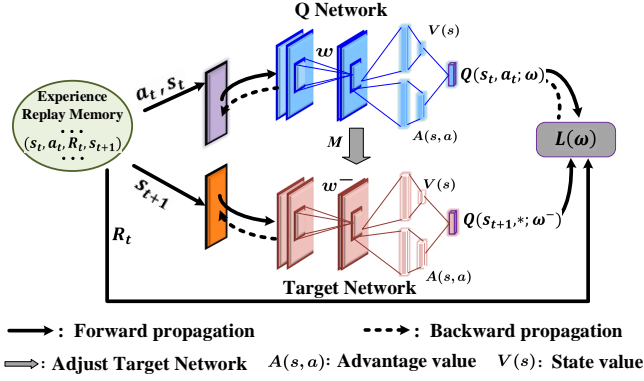

Figure 6: The architecture of D3QN

$$w^- = (1 - \tau) \cdot w^- + \tau \cdot w \qquad (25)$$

Here, $\tau$ is the update rate of the weights $w^-$. The employment of soft updates not only helps strike a balance between exploration and exploitation of the learning, but also enhances stability and facilitates experience accumulation.

The flowchart of the D3QN-based optimization approach for microgrid control can be illustrated in Figure 7, while the implementation steps are depicted in Algorithm 1. Overall, the learning process is carried out over a number of episodes. In each episode, a total of $K$ time series sequences are first sampled from the whole dataset and used as inputs to the proposed approach. Here, each time series is made up of a sequence of $3(T+1)$ external inputs plus an internal variable to represent the state in (21) and (22). By simulating the microgrid system one step forward, each time series will produce a tuple $(s'_t, a'_t, R'_t, s'_{t+1})$, where $a'_t$ can be obtained using the $\epsilon$-greedy algorithm with the optimal policy $\pi^*$. The system can thereby receive an immediate reward $R'_t$ (23) and move to the next state of charge $E'_{b,t+1}$ (9). This tuple will be stored in the experience replay memory for later use. Subsequently, a mini-batch of tuple $(s_t, a_t, R_t, s_{t+1})$ will be randomly sampled from the experience replay memory to update the optimal policy $\pi^*$ using the D3QN network as described above. The updated $\pi^*$ will be then used in microgrid simulation to decide the action of the next time series. The learning continues until a maximum number of episodes is reached, where the optimal Q network is obtained. In applying the control actions, this paper implemented the $\epsilon$-greedy strategy with a decreasing $\epsilon$ using an exponential decay factor of 0.999. The minimum value of $\epsilon$ was truncated to 0.01 ($\epsilon$ was also fixed as 0.01 for the application of the optimized Q network). This enables the agent to freely explore the microgrid environment at the beginning and then gradually rely on the learned knowledge while still having the chance to further improve the obtained policy in question.

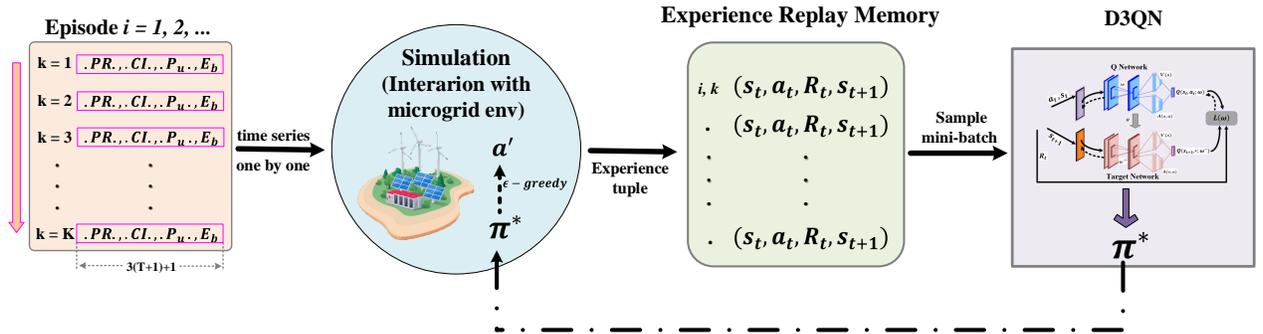

Figure 7: Flowchart of the D3QN-based approach

---

**Algorithm 1:** D3QN-based Approach for Microgrid Control

**Input**     : predicted future or historical data - $PR, CI, P_u, E_b$
**Output**    : optimal policy $\pi^*$ (Q network)
**Procedure:**

1. Initialize microgrid configurations and experience replay memory;
2. Initialize parameters and network weights $(w^-, w)$;
3. **for** $i = 1$ to $Episode_{max}$ **do**
4.     Randomly initialize $E_{b,t}$ within $[E_b^{min}, E_b^{max}]$;
5.     Sample a sequence containing K time series from the whole dataset;
6.     **for** $k = 1$ to $K$ **do**
7.         Observe the state $s'_t$ based on (21) or (22) ;
8.         Select and execute an action $a'_t$ using Q Network based on $\epsilon$- greedy;
9.         Calculate reward $R'_t$ using (23) and $E_{b,t+1}$ using (9);
10.        Observe the next state $s'_{t+1}$;
11.        Store the tuple $(s'_t, a'_t, R'_t, s'_{t+1})$ in the Experience Replay Memory;
12.        Randomly sample a mini-batch of $(s_t, a_t, R_t, s_{t+1})$ from Replay Memory;
13.        Estimate the $Q(s_t, a_t; w)$ using Q Network with $s_t$ and $a_t$;
14.        Calculate the TD target using Target Network with $s_{t+1}$ and $R_t$;
15.        Calculate loss $L(w)$ using (16) and (17);
16.        Optimize the weights $w$;
17.        **if** $k$ mod $M = 0$ **then**
18.             Soft update the weights $w^-$ of Target Network using (25);

# 5. SIMULATIONS AND PERFORMANCE EVALUATION

*5.1. Microgrid Description and Parameter Configuration*

The proposed approach was evaluated using an open-source dataset for an integrated district energy system in US [36], which contains hourly load, electricity price and PV generation for 12 months (with a time interval of $\triangle t = 1h$). In addition, the hourly carbon intensity data was collected from the WattTime API [37]. The WattTime API offers access to carbon intensities of electrical grids in different regions of the world. Some pre-processing tasks (such as scaling and cleaning) were also performed on the integrated dataset, and the completed dataset was made accessible on Github [38].

Then, the time horizon of the PB and PF schemes was set as $T = 24$. In other words, this paper treats the microgrid control problem as using the data of the future/past 24 time steps to determine the charging or discharging actions. Generally, the battery capacity $E_b^c$ in a microgrid is designed to be much smaller than the peak unmet due to the high capital and maintenance costs, while the charging/discharging rate should ensure that the battery can theoretically be fully charged or discharged within a limited time [15, 39]. Moreover, lithium-ion batteries may fail when overcharged or over-discharged, and can even explode in extreme cases [40]. To maintain the security of the microgrid, the maximum ($E_b^{max}$) and minimum ($E_b^{min}$) states of charge are thus defined as (26). Given these, Table 1 lists the configurations of the microgrid considered in this paper.

$$\begin{cases} E_b^{max} = 0.9 \cdot E_b^c \\ E_b^{min} = 0.1 \cdot E_b^c \end{cases} \quad (26)$$

Table 1: Configuration of the microgrid

| Item | Value | Unit | Item | Value | Unit | Item | Value | Unit | Item | Value | Unit |
|---|---|---|---|---|---|---|---|---|---|---|---|
| $E_b^c$ | 1000 | kWh | $E_{max}$ | 300 | kWh | $\triangle t$ | 1 | h | $T$ | 24 | —— |
| $E_b^{max}$ | 900 | kWh | $\eta_b$ | 0.95 | —— | $E_b^{min}$ | 100 | kWh | $P_g^{max}$ | 4450 | kW |

In the proposed D3QN, the same Q-network structure (excluding input and output layers) was adopted for both the value function and advantage function streams. The structure and parameter settings of the Q network designed in this paper are shown in Table 2. The length of time series sequences in each episode was set as $K = 168$ being the number of hours in a week. In other words, we train the D3QN on a week's worth of samples in each episode, updating the Q Network for each sample (hour). Moreover, by trial and error, the parameters in the reward function (23) were set as: $\alpha = 0.25, \beta = 1.5, \lambda = 0.02$.

To fully evaluate the effectiveness of the proposed D3QN approach, this paper provides a comprehensive comparison between vanilla (traditional) DQN and D3QN, PB and PF schemes, as well as hard and soft updates. In addition, the vanilla DQN (hard update) together with the common control scheme (where only the current data $(PR_t; CI_t; P_{u,t})$ is used to compose the state space for inferring actions from a policy) was adopted as the baseline to examine the superiority of the proposed approach. The implementation was coded in Python using the Keras library and was run on Google Colab with high RAM and no hardware accelerators.

Table 2: Structure and parameter settings of the network (each stream)

| Item | Value | Item | Value |
|---|---|---|---|
| no. of FC layers | 3 | batch size | 64 |
| no. of neurons per FC layer | 64 | learning rate | 0.0003 |
| update frequency of Target Network: $M$ | 24 | discount factor: $\gamma$ | 0.99 |
| length of time series sequences: $K$ | 168 | random exploration: $\epsilon$ | $0.3 \to 0.01$ |
| updating rate of Target Network: $\tau$ | 0.01 | activation function | Relu |

*5.2. Convergence Evaluation*

The convergence of the proposed microgrid control approach is first evaluated in this subsection. The convergence curves of vanilla DQN and D3QN under different implementations are shown in Figure 8. To visualise the learning process more clearly, the cumulative reward for each episode (scaled down by 1000 times) is the average of the past 50 episodes. Due to the fact that each episode is only equivalent to a period of consuming 168-hour data (one week) and the random exploration $\epsilon$ was eventually truncated to a minimum value of 0.01, it is reasonable to see the converged reward fluctuates within a certain range. Table 3 presents the average and variance of the converged reward of the 10 approaches during 10,000-12,000 episodes. Under the same or similar learning conditions, the following can thus be observed: 1) PB and PF schemes (whether in the vanilla or D3QN architectures) both substantially outperformed the common data-driven scheme, indicating the validity of bringing either historical or predicted data of the external variables into the state space. The control actions can thereby be more proactive to future or past changes of the microgrid system rather than only reactive to its current statuses. 2) The PB scheme was more likely to achieve higher rewards than the PF scheme, either using vanilla or D3QN architecture. In addition, although the structure of D3QN is more complex than vanilla's, it offered much greater rewards. 3) Compared with hard update, although soft update progressed slower in the initial learning stage, it eventually reached convergence almost at the same time as the former and obtained a higher reward.

Table 3: Converged rewards (average and variance) from different microgrid control approaches

| Approach | Variance (Reward) | Average (Reward) | Approach | Variance (Reward) | Average (Reward) |
|---|---|---|---|---|---|
| Vanilla-PB-soft | 0.00027 | 0.8509 | **D3QN-PB-soft** | **0.00047** | **0.9392** |
| Vanilla-PB-hard | 0.00043 | 0.6639 | D3QN-PB-hard | 0.00055 | 0.9158 |
| Vanilla-PF-soft | 0.00024 | 0.6549 | D3QN-PF-soft | 0.00050 | 0.8109 |
| Vanilla-PF-hard | 0.00048 | 0.6215 | D3QN-PF-hard | 0.00029 | 0.6752 |
| Vanilla-common-hard | 0.00023 | -0.0155 | D3QN-common-hard | 0.00020 | 0.0753 |

*5.3. Cumulative Reward Evaluation*

To further validate the effectiveness of the proposed approach, the cumulative reward over the whole operational period (i.e., one year given this dataset) is also evaluated on both actual values and simulated predictions. Specifically, as the PF scheme does not rely on predictions, we only tested PF scheme with actual values. In contrast, the PB scheme was tested on both actual values (i.e., assuming 100% prediction accuracy) and simulated predictions. Here, the simulated predictions of external variables were generated by adding Gaussian noises to the actual values, with a mean of 0 and a variance equal to 10% of the actual data. This was to introduce randomness and uncertainty to simulate the predicted future values. The simulation results against the actual and predicted values are shown in Figure 9 and Figure 10, respectively. It is worth noting that results from using actual values for the PF scheme were kept in Figure 10 for the convenience of comparisons. Additionally, Table 4 gives the cumulative rewards received in both actual and predicted cases.

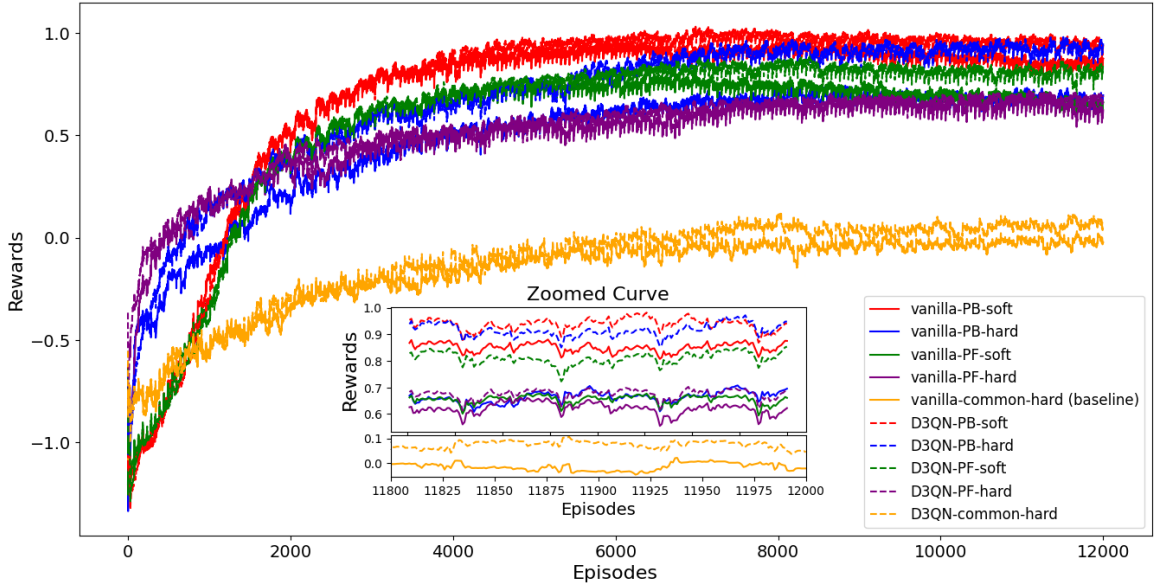

Figure 8: The evolution of rewards during the training stage

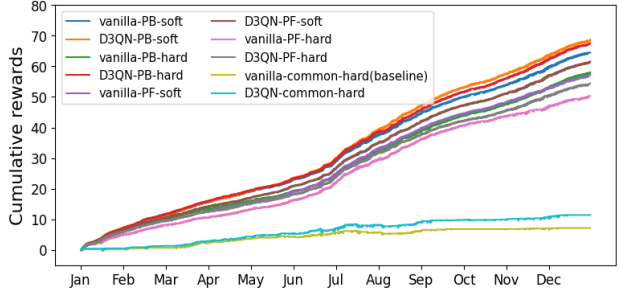

Figure 9: Experimental results using the actual values of external variables

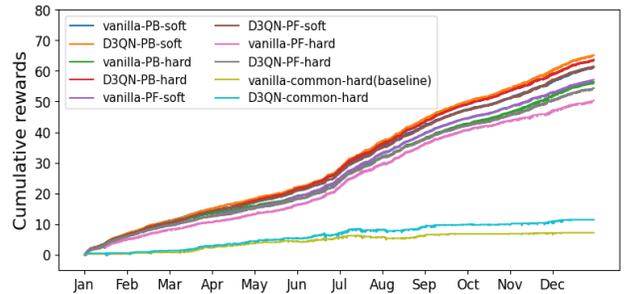

Figure 10: Experimental results using the simulated predictions of external variables

Clearly, the D3QN-PB-soft approach achieved the best results, even when using predicted values with 10% noise. This is consistent with the conclusions drawn in Section 5.2 (i.e., PB, D3QN and soft update were all in favour of producing higher cumulative rewards). Additionally, both PB and D3QN tended to have more improvement upon the policy learning than soft update. In detail, the D3QN-PB-soft approach produced a cumulative reward that is 1.8% (using the actual as predictions) and 2.2 % (using noise data as predictions) higher than that of D3QN-PB-hard, showing a slight improvement of soft update. In contrast, the D3QN-PB-soft approach gained 11.7% (actual) and 5.9% (noise) higher cumulative rewards than that of D3QN-PF-soft, and 6.6% (actual) and 6.5% (noise) higher cumulative rewards than that of vanilla-PB-soft, which demonstrated remarkable improvements of PB and D3QN.

Table 4: Cumulative rewards from using actual and predicted data

| Approach | Actual | Predicted | Approach | Actual | Predicted |
| --- | --- | --- | --- | --- | --- |
| Vanilla-PB-soft | 64.50 | 61.21 | **D3QN-PB-soft** | **68.74** | **65.16** |
| Vanilla-PB-hard | 58.05 | 56.40 | D3QN-PB-hard | 67.52 | 63.77 |
| Vanilla-PF-soft | 57.10 | —— | D3QN-PF-soft | 61.52 | —— |
| Vanilla-PF-hard | 50.31 | —— | D3QN-PF-hard | 54.39 | —— |
| Vanilla-common-hard | 7.13 | -0.0155 | D3QN-common-hard | 11.37 | 0.0753 |

*5.4. Resilience Evaluation and Comparison with other methods*

Table 4 also reveals that the performance of the PB scheme declined (across all approaches) when noise was introduced to the predictions of external variables. We here further evaluate PB's resilience to these prediction errors, specifically under what error conditions the PB scheme outperforms PF, and under what circumstances the opposite occurs. In addition to the 10% Gaussian noise added before, we also studied the situations where predictions came with 5%, 12.5% and 15% errors, respectively, as shown in Figure 11. The results indicate that the PB scheme outperforms the PF scheme when the prediction error is within 12.5%. However, for errors exceeding 12.5%, the PF scheme becomes more advantageous not only because it has a higher return, but also because it avoids the need to design additional predictive models.

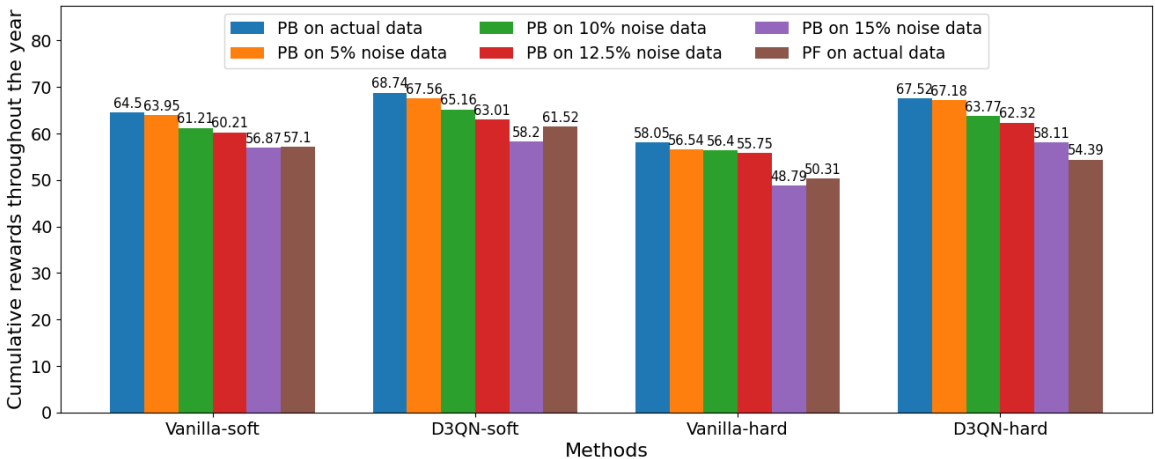

Figure 11: Comparative results between using the actual and noisy data at different levels

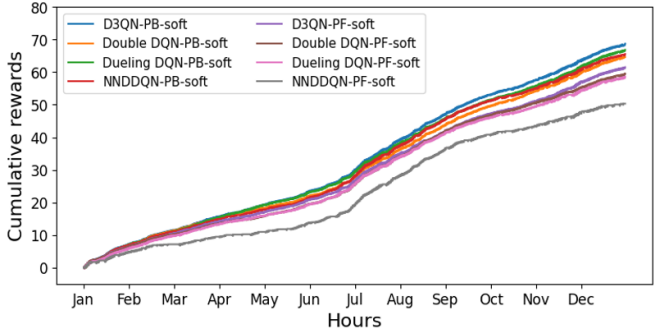
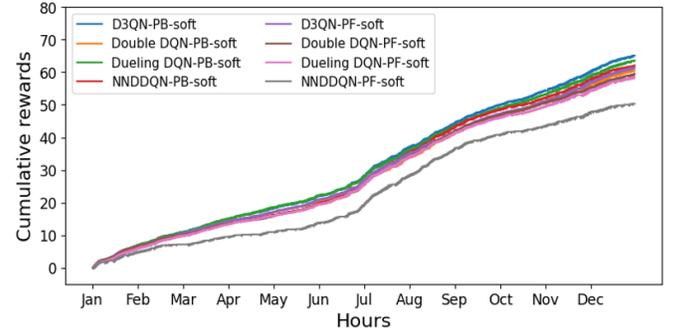

Figure 12: Comparison with other DRL methods using actual values

Figure 13: Comparison with other DRL methods using predicted values

In previous subsections, we have discussed the superiority of our approach compared to traditional DQN (vanilla DQN). Here, we further compared the D3QN-based optimization approach with other popular DRL-based optimization methods, including DDQN, Duling DQN, NN-DDQN. These methods have been widely used to solve similar microgrid control problems and achieved remarkable results [14, 15, 41]. Figure 12 and Figure 13 show the comparison of annual cumulative rewards using actual and predicted values, respectively. The annual cumulative rewards are listed in Table 5. Here, it can be observed that our D3QN approach is much better than the other three methods no matter using the actual values or the predicted values (10% noise). This holds truth for both the PB and PF schemes.

Table 5: Annual cumulative rewards of different methods

| Approach | Actual | Predicted | Approach | Actual | Predicted |
| --- | --- | --- | --- | --- | --- |
| **D3QN-PB-soft** | **68.74** | **65.16** | **D3QN-PF-soft** | **61.52** | —— |
| Double DQN-PB-soft | 64.67 | 60.55 | Double DQN-PF-soft | 59.54 | —— |
| Dueling DQN-PB-soft | 66.85 | 63.62 | Dueling DQN-PF-soft | 58.63 | —— |
| NN DDQN-PB-soft | 65.56 | 62.13 | NN DDQN-PF-soft | 50.31 | —— |

### 5.5. Objectives Evaluation

This subsection is to assess each individual objective within the reward function, i.e., operational costs, carbon emissions, battery degradation and peak load. We took the D3QN-PB-soft approach to demonstrate this objective evaluation as it produced the best rewards in various scenarios presented previously. Figure 14 - Figure 17 show the cumulative values of each objective over the year from using actual and predicted data (10% noise). The results without considering the respective objective in the proposed approaches are also presented there for comparison.

It can be seen that the inclusion of each objective in the reward function all played a positive role in controlling the operational cost, carbon emission, battery degradation and peak load of the microgrid. Specifically, the proposed D3QN-PB-soft approach using a holistic reward function reduced the operational costs by 7514.61% and carbon emissions by 133.90% on the actual data (6020.75% and 342.58%, respectively, on the predicted data) - see Figure 14 and 15. As also shown in Figure 16 and 17, the proposed approach successfully reduced the battery degradation and peak load. In addition, some interesting observations can be drawn between different objectives and seasons. For example, the carbon objective exhibited higher sensitivity during the winter, whereas the operational cost showed greater sensitivity during the summer. The peak load exceeded the limit only during specific periods in the summer, while battery degradation was relatively insensitive to seasonal variations.

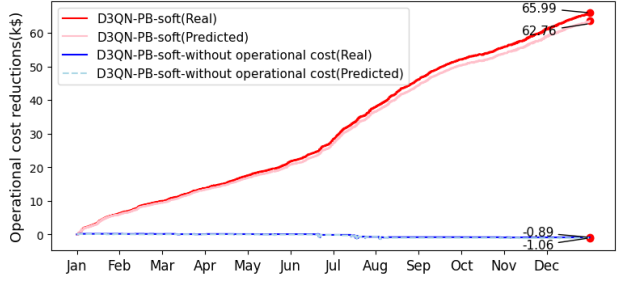

Figure 14: Cumulative operational cost reductions (on a real-world scale) using actual and predicted data

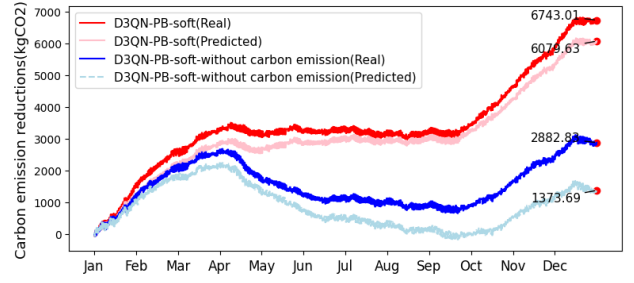

Figure 15: Cumulative carbon emission reductions (on a real-world scale) using actual and predicted data

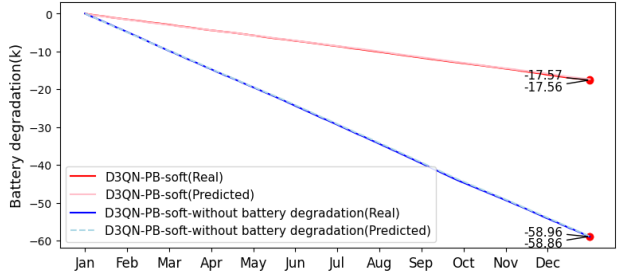

Figure 16: Cumulative battery degradations (on a real-world scale) using actual and predicted data

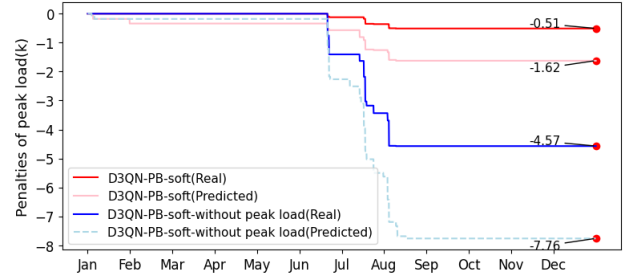

Figure 17: Cumulative penalties of peak load (on a real-world scale) using actual and predicted data

Table 6: Annual cumulative values of each objective (on a real-world scale) regarding different weight settings

| $(\alpha, \beta, \lambda)$ | Actual | | | | Predicted | | | |
| --- | --- | --- | --- | --- | --- | --- | --- | --- |
| | CCB(kgCO$_2$) | CPP(k) | CBD(k) | CMP(k$) | CCB(kgCO$_2$) | CPP(k) | CBD(k) | CMP(k$) |
| (**0.25**,1.5,0.02) | **6743.01** | **-0.51** | **-17.56** | **65.99** | **6079.63** | **-1.62** | **-17.57** | **62.76** |
| (**0.125**,1.5,0.02) | **5318.77** | -0.73 | -16.42 | 66.78 | **5015.66** | -1.95 | -16.34 | 66.72 |
| (**0.5**,1.5,0.02) | **8223.54** | -0.75 | -16.01 | 64.41 | **8095.99** | -1.79 | -15.7 | 63.7 |
| (0.25,**1.25**,0.02) | 6045.68 | **-0.91** | -16.94 | 65.95 | 6223.77 | **-1.9** | -16.68 | 66.08 |
| (0.25,**2**,0.02) | 6262.67 | **-0.41** | -16.19 | 66.38 | 6154.54 | **-1.41** | -15.69 | 66.65 |
| (0.25,1.5,**0.01**) | 5931.37 | -1.18 | **-19.07** | 68.14 | 5976.54 | -2.36 | **-18.48** | 68.53 |
| (0.25,1.5,**0.04**) | 6611.37 | -0.59 | **-12.64** | 61.06 | 6543.07 | -1.76 | **-12.35** | 60.3 |

We also evaluated the impact of different weight values in the reward function (23) on the four objectives. Apart from the parameter settings mentioned earlier (i.e., $\alpha = 0.25, \beta = 1.5, \lambda = 0.02$), we also tested six more settings, where we either increased or decreased the values of $\alpha, \beta$ or $\lambda$ while keeping other parameters constant. The results are shown in Table 6 (using D3QN-PB-soft), where CBD, CPP, CMP and CCB represent the annual cumulative battery degradation, penalty of peak load and operational cost reductions and carbon emission reductions, respectively. It is evident that the importance of each objective can be controlled by adjusting the corresponding weight parameter. For example, changing $\alpha$ from 0.25 to 0.5 led to a 21.96% increase in CCB (from 6743.01 to 8223.54 kgCO$_2$). Conversely, reducing $\alpha$ from 0.25 to 0.125 resulted in a 21.12% decrease in CCB. Moreover, the improvement of one objective often comes at the expense of others, indicating the conflicting relationship between the four objectives. This allows decision-makers to emphasize more on a specific objective as needed by assigning a smaller value to the corresponding weight parameter.

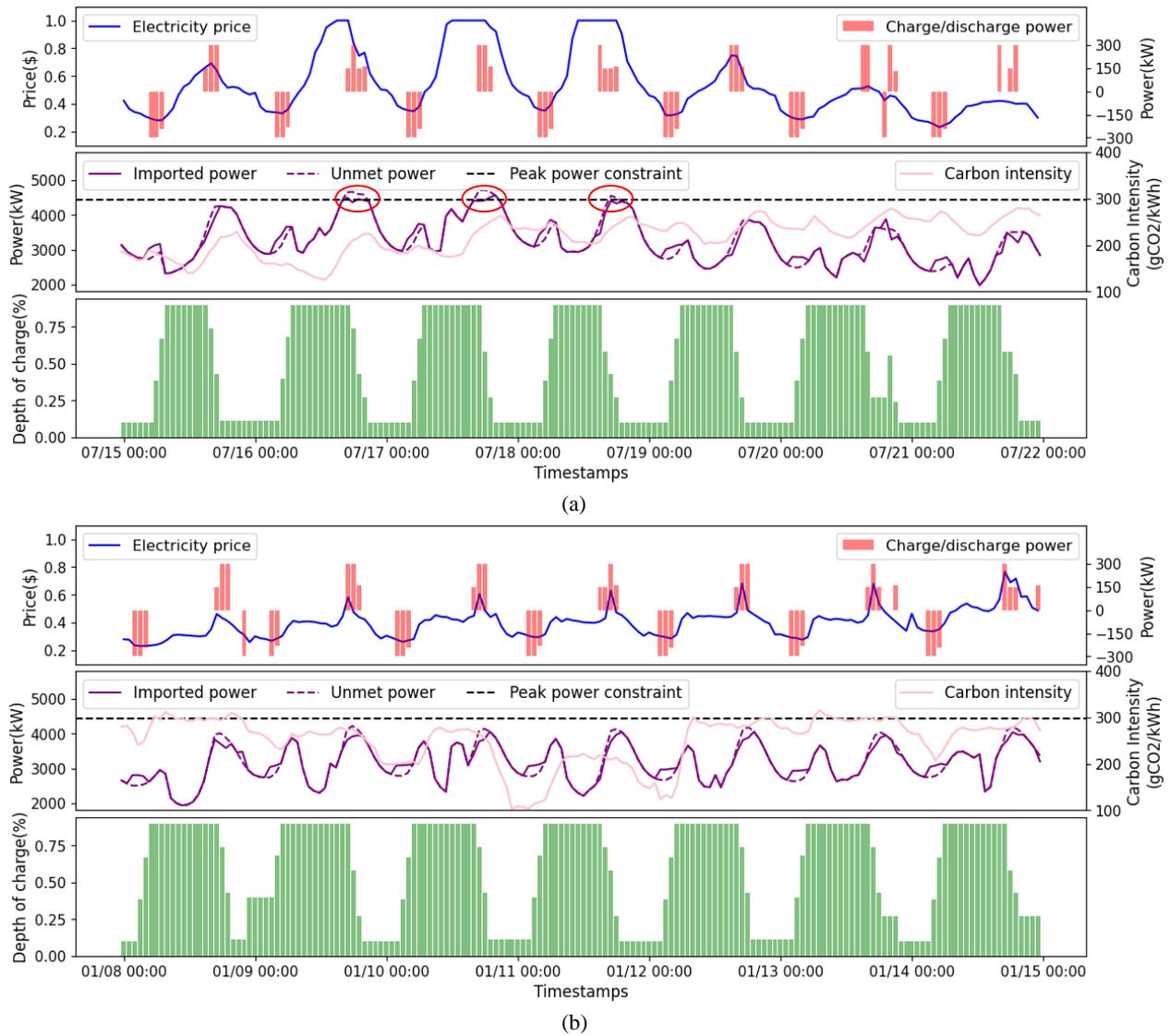

Figure 18: A week-long visualization of battery behaviors in summer (a) and winter (b)

## 5.6. Simulation Visualization

To further demonstrate the effectiveness of the proposed approach (again taking the example of D3QN-PB-soft), Figure 18 shows the resultant charging and discharging behaviour of the battery (a week in summer and winter, respectively). It can be seen that the control strategy was not only suitable for the winter - see Figure 18 (b), but can also be effective in the summer when the variation of electricity price and unmet power were quite high - see Figure 18 (a). This means that the proposed approach can effectively guide the agent to charge during low price and carbon intensity periods as well as to discharge during high price and carbon intensity periods. When the unmet power exceeded the peak constraint in summer, the agent can promptly control the battery to adopt a discharge strategy to ease the peak load pressure - see the peak hours between 07/16 and 07/18 (marked by red circles). In addition, as the charging and discharging operations need to meet the battery's constraints (8), some charging behaviors in addition to the five specified ones reasonably appeared in the simulation.

## 5.7. Evaluation with an Additional Dataset

To verify whether the proposed approach remains effective on different data, we conducted further simulations by altering the original dataset. Here, unmet power $P_{u,t}^n$, carbon intensity $CI_t^n$ and electricity price $PR_t^n$ in the new dataset were produced using (27)-(29), respectively. The original microgrid variables exhibit periodic fluctuations. Consequently, after being multiplied by a sine (sin) or cosine (cos) function, the resulting data continues to demonstrate periodic behavior [42], as shown in Figure 19. This ensures that the new dataset also aligns with the periodic variation patterns observed during Microgrid operations. Except for $P_g^{max}$ being reset to 6000, all other parameter settings remain the same as specified in Section 5.1.

$$P_{u,t}^n = P_{u,t}(0.5 + |sin(3t)|) \tag{27}$$

$$CI_t^n = CI_t(0.5 + |cos(3t)|) \tag{28}$$

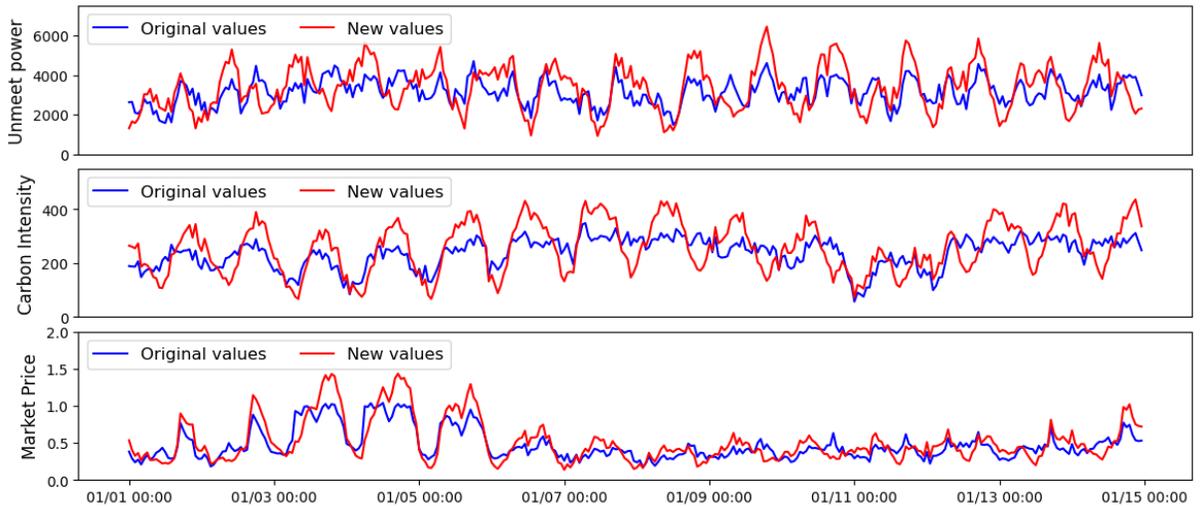

Figure 19: An illustrative example of the new and original datasets

$$PR_t^n = PR_t(0.5 + |cos(4t)|) \tag{29}$$

As a result of the new dataset, the annual cumulative rewards from the proposed approach, along with other DRL methods, are presented in Table 7. As in previous experiments, the rewards were scaled down by a factor of 1000. It is evident that the D3QN-PB-soft continues to outperform the others, both on actual and predicted (with 10% added noise) data. Additionally, Figure 20 illustrates a week-long simulation using the D3QN-PB-soft on the new dataset, showing similar performance as seen from the original dataset. Finally, it led to 37.02 t$CO_2$ CCB, -0.55K CPP, -22.39k CBD and 104.74k$ CMP on actual data, while achieving 23.68 t$CO_2$ CCB, -0.50K CPP, -24.87k CBD and 79.68k$ CMP on predicted data. These results, once again, confirm the effectiveness of our approach.

Table 7: Annual cumulative rewards of the new dataset using different methods

| Approach | Actual | Predicted | Approach | Actual | Predicted |
| --- | --- | --- | --- | --- | --- |
| **D3QN-PB-soft** | **104.65** | **68.67** | D3QN-PF-soft | **100.50** | —— |
| Double DQN-PB-soft | 102.12 | 64.83 | Double DQN-PF-soft | 97.96 | —— |
| Dueling DQN-PB-soft | 98.49 | 66.54 | Dueling DQN-PF-soft | 93.18 | —— |
| NN DDQN-PB-soft | 98.84 | 14.65 | NN DDQN-PF-soft | 76.59 | —— |
| Vanilla-PB-soft | 94.57 | 65.31 | Vanilla-PF-soft | 93.01 | —— |

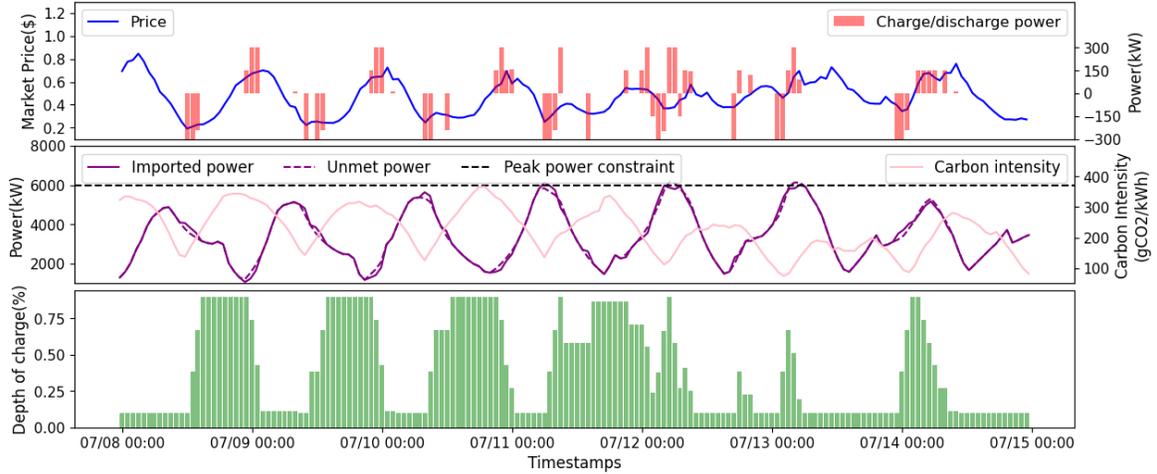

Figure 20: A week-long simulation on the new dataset

## 6. Conclusion

This paper presented a holistic data-driven power optimization approach based on D3QN to manage the operational control of a microgrid system integrated with RES and ESS. The proposed approach was able to jointly handle economic, environmental and infrastructural considerations, plus proactively accounting for uncertainties from both the supply and demand sides. To achieve that, an inclusive reward function was designed to incorporate operational costs, carbon emissions, peak load and battery degradation. Two data-driven control schemes, i.e., prediction-based (PB) and prediction-free (PF), were adopted based on different problem formulations under the proposed D3QN architecture for the optimization of battery charging and discharging behaviours. The findings from extensive simulations not only validated the effectiveness of the proposed approach but also offered guidance for its practical implementations. The contributions of this research can significantly advance the development of sustainable and resilient microgrid systems with high renewable penetration both onsite and from the grid. In the future work, we will implement the proposed approach on microgrid systems integrated with multiple RES (e.g., photovoltaic and wind) and ESS (e.g., battery and thermal storage) to maximise its potential in bringing clean and affordable energy.

## Acknowledgment

Fulong Yao acknowledges the financial support of China Scholarship Council and Newcastle University for his PhD study. This work was supported in part by Royal Society and NSFC under grants IEC\NSFC\201107 and 62111530154.